\documentclass[aps,prl,reprint,groupedaddress,showpacs,amsmath,amssymb,twocolumn,floatfix]{revtex4-1}
\usepackage{graphicx}
\usepackage{bm}
\usepackage{color}

\begin{document}

\title{The Superfluid Glass Phase of $^3$He\,-\,A}

\author{J.I.A. Li}
\email[]{jiali2015@u.northwestern.edu}
\author{J. Pollanen}
\author{A.M. Zimmerman}
\author{C.A. Collett}
\author{W.J. Gannon}
\author{W.P. Halperin}
\email[]{w-halperin@northwestern.edu}
\affiliation{Northwestern University, Evanston, IL 60208, USA}

\date{\today}



\maketitle

It is established theoretically that an ordered state with continuous symmetry is inherently unstable to arbitrarily small amounts of disorder~\cite{Lar.70,Imr.75}. This principle is of central importance in a wide variety of condensed systems including superconducting vortices~\cite{Gin.96,Emi.99}, Ising spin models~\cite{Sim.11} and their dynamics~\cite{Fis.01}, and liquid crystals in porous media~\cite{Rad.99, Bel.00}, where some degree of disorder is ubiquitous, although its experimental observation has been elusive.  Based on these ideas it was predicted~\cite{Vol.96} that $^3$He in high porosity aerogel would become a superfluid glass. We report here  our nuclear magnetic resonance measurements on $^3$He in aerogel demonstrating  destruction of  long range orientational order of the intrinsic superfluid orbital angular momentum, confirming the existence of a superfluid glass.  In contrast,  $^3$He-A generated by warming from superfluid $^3$He-B has perfect long-range orientational order providing a  mechanism for switching off this effect.

\begin{figure}[htbp]
\centerline{\includegraphics[height=0.35\textheight]{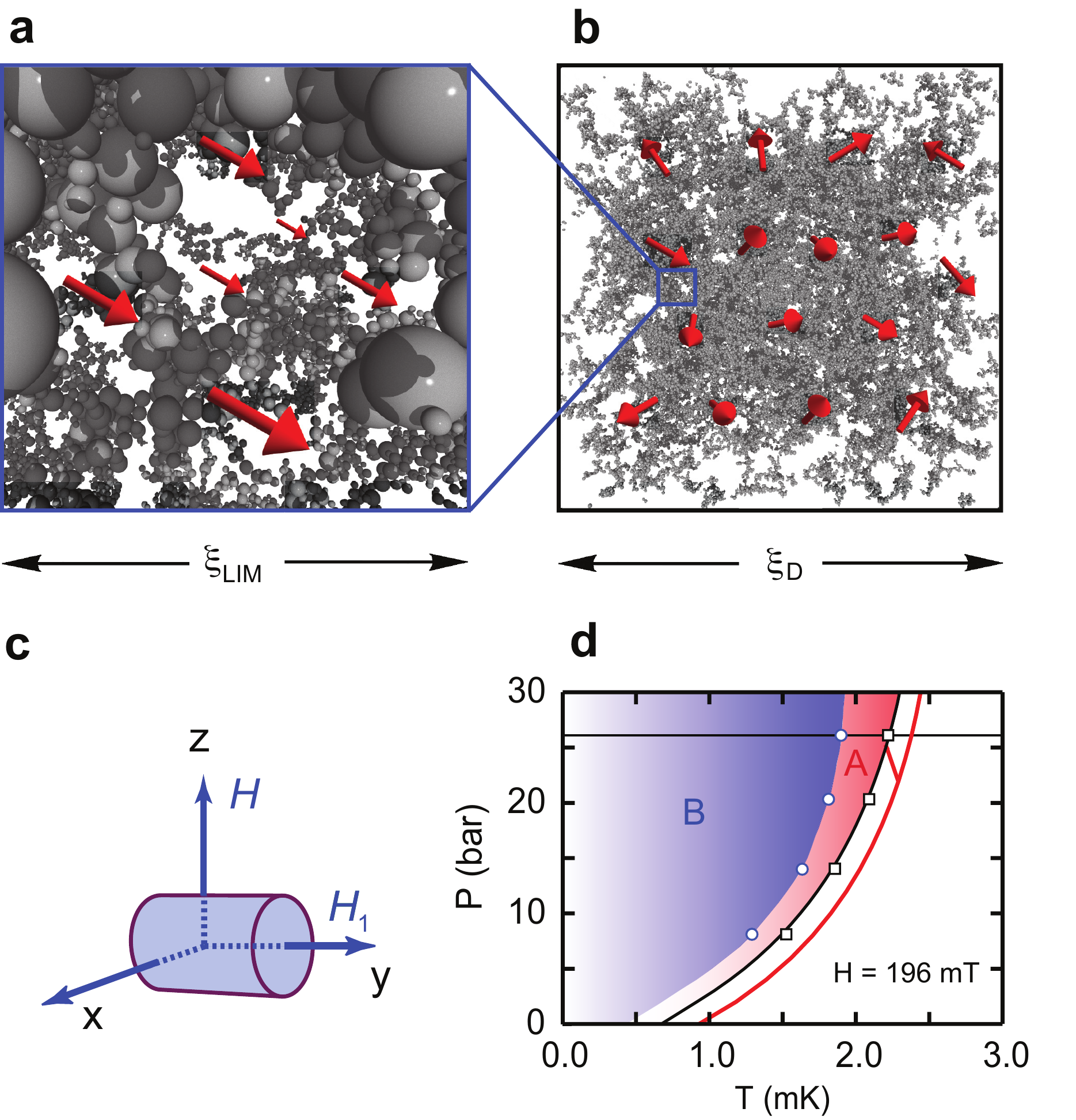}}
\caption{\label{fig1}(Color online). {\bf{Superfluid $^3$He in aerogel.}} a, b) Numerical simulation of a 98\% aerogel~\cite{Pol.12a} using a diffusion limited cluster aggregation algorithm.  Red arrows represent the orbital angular momentum distribution for a LIM state, uniform on a short length scale, $\xi_{LIM}$, called the LIM length, but disordered on a length scale smaller than the dipole length, $\xi_D \sim 8\, \mu$m. c) Geometry of the experimental setup. The cylinder axis of the aerogel sample and the RF field $\bm{H}_1$ are both along the $y$-axis while the external magnetic field, $\bm{H}$, is oriented along the vertical $z$-axis; d) Pressure versus temperature phase diagram for superfluid $^3$He for $H = 196$  mT imbibed in an isotropic aerogel~\cite{Pol.11}, the same as for the present work. The open squares are the superfluid phase transitions and the open circles are the B to A-phase transitions on warming. The solid red curves in the background correspond to $T_c$ and $T_{BA}$ for pure $^3$He in zero magnetic field.}
\end{figure}
\begin{figure}
\centerline{\includegraphics[width=0.35\textheight]{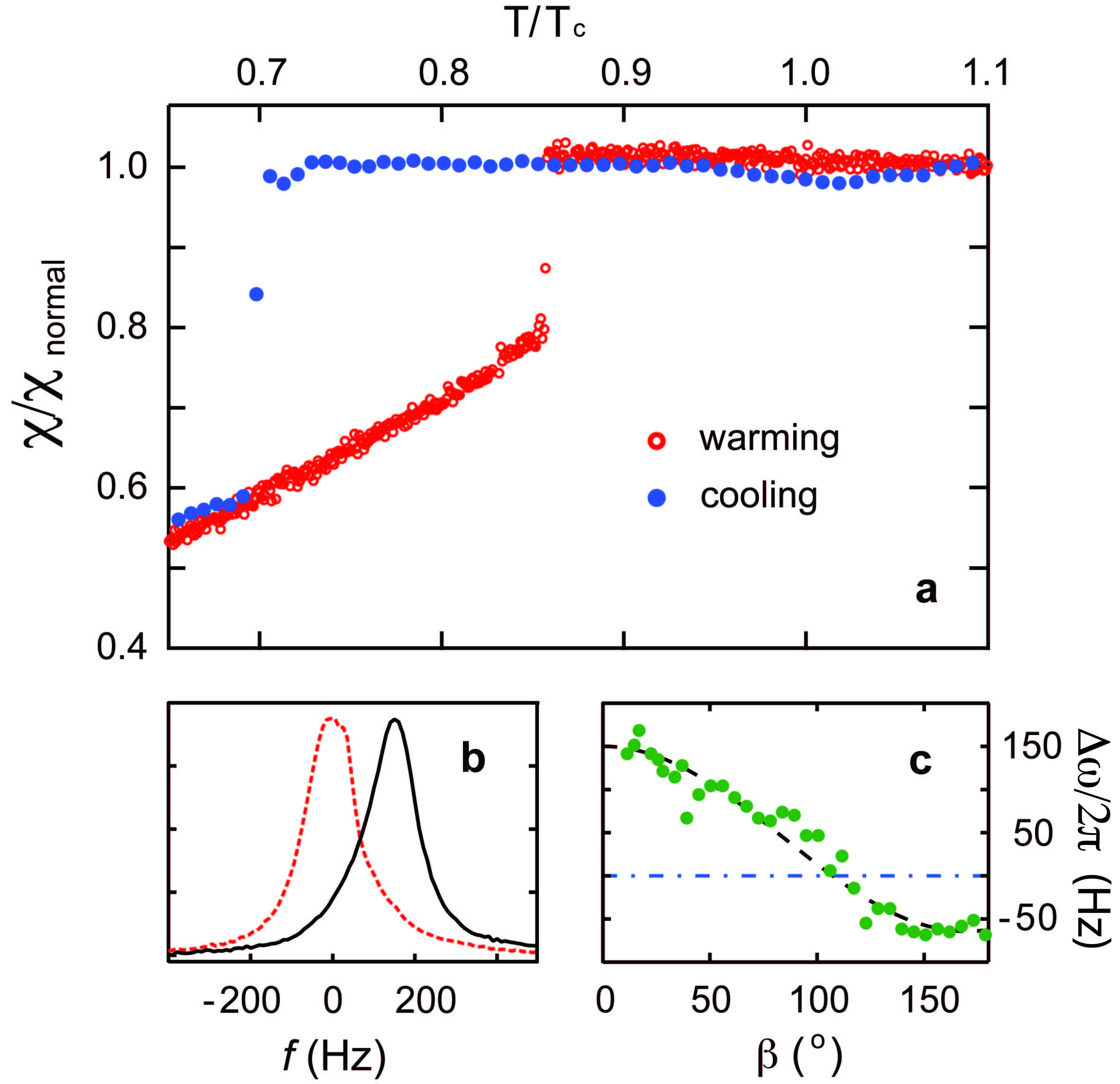}}
\caption{\label{fig2} (Color online). {\bf{Identification of the superfluid state.}} a) Liquid susceptibility relative to the normal state on warming (open red circles) and cooling (closed blue circles) versus reduced temperature in $H=196$ mT. The data are spaced at  $\approx 2 \,\,\mu$K intervals on warming and $\approx 15 \,\,\mu$K on cooling. The liquid susceptibility was obtained after subtracting the  contribution from the several layers of solid $^3$He adsorbed to the surface of aerogel, well-known to have a Curie-Weiss temperature dependence~\cite{Spr.95,Bra.10,Pol.11}.  The jump in susceptibility on cooling or warming marks first order transitions, $T_{AB}$ ($T_{BA }$). The transition temperature $T_c$ is determined as the point of onset of NMR frequency shifts. b) NMR spectra for $^3$He in aerogel in the normal state (red dashed curve) and the A-phase (black curve), obtained in $H=196$ mT and $P=26.3$ bar on warming from the B-phase at $T/T_c =0.85$. c) Frequency shift versus NMR tip angle $\beta$ for the A-phase nucleated from the B-phase in $H=196$ mT (green circles). The dash-dot blue line is the bare Larmor frequency, $\omega_L$, while the dashed black curve is the fit to the theory~\cite{Bri.75} for the dipole ordered axial state.}
\end{figure}
Close to the absolute zero of temperature, liquid $^3$He  condenses into  a $p$-wave superfluid of Cooper pairs resulting in two phases with fundamentally different symmetry:  the isotropic B-phase and an anisotropic A-phase.  In zero magnetic field $^3$He-A appears  in a small corner of the pressure versus temperature phase diagram shown in Fig.~1(d).   Its anisotropy, a paradigm for more recently discovered unconventional superconductors \cite{Nor.13},  is characterized by the orientation of its order parameter  defined by orbital angular momentum and spin induced by magnetic field, $\bm{\hat{l}}$ and $\bm{\hat{s}}$.  The spin is necessarily parallel to an applied magnetic field, $\bm{H}$; however,  the orbital angular momentum has  continuous  rotational symmetry.  That symmetry can be broken, for example at a wall or interface to which  $\bm{\hat{l}}$ must be perpendicular, thereby defining a preferred direction on a macroscopic scale.  
 
 \begin{figure}
\centerline{\includegraphics[height=0.4\textheight]{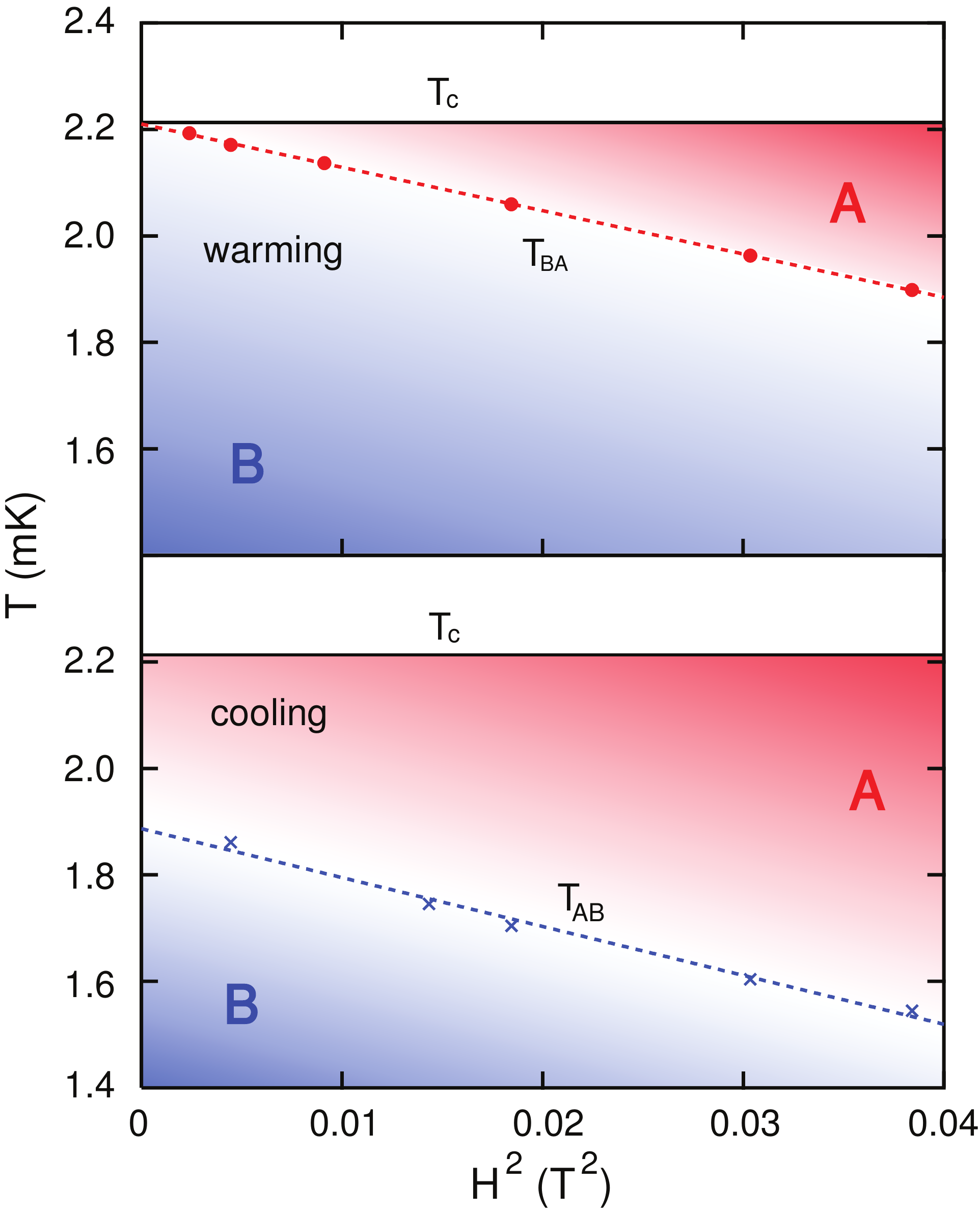}}
\caption{\label{fig3}(Color online). {\bf{Superfluid phase diagrams.}} Superfluid phase diagram $T(H^2)$ for $^3$He at a pressure of $P=26.3$ bar,  for warming from the B-phase (upper panel) contrasted with cooling from the normal fluid (lower panel). The solid black line is $T_{c} = 2.213$ mK, independent of the magnetic field. The red circles are $T_{BA}$ corresponding to B-to-A phase transitions on warming; the blue crosses are $T_{AB}$, the A-to-B phase transition after supercooling uniformly by $\approx 300\,\,\mu$K. The precise parallel behavior of $T_{AB}(H^2)$ and $T_{BA}(H^2)$ indicates that the two phases,  formed by warming from the B-phase and by cooling from the normal fluid, are the same  state.}
\end{figure}

Volovik proposed~\cite{Vol.96} that this long range orientational coherence of angular momentum would be destroyed by random microscopic disorder that can be realized if the  $^3$He is imbibed in highly porous silica aerogel as shown in our simulation Fig.~1(a,b).   This sensitivity to small amounts of disorder on a microscopic scale was  discussed by  Larkin~\cite{Lar.70} and Imry and Ma~\cite{Imr.75} for a broad range of physical phenomena~\cite{Gin.96,Emi.99,Sim.11,Fis.01,Rad.99, Bel.00} and which we refer to as the LIM effect.  If this proposal is correct then in the LIM state the order parameter structure of the superfluid will be completely hidden, a behavior of potential significance for understanding  exotic superconductors such as URu$_2$Si$_2$~\cite{Myd.11}.  

\begin{figure}
\centerline{\includegraphics[height=0.40\textheight]{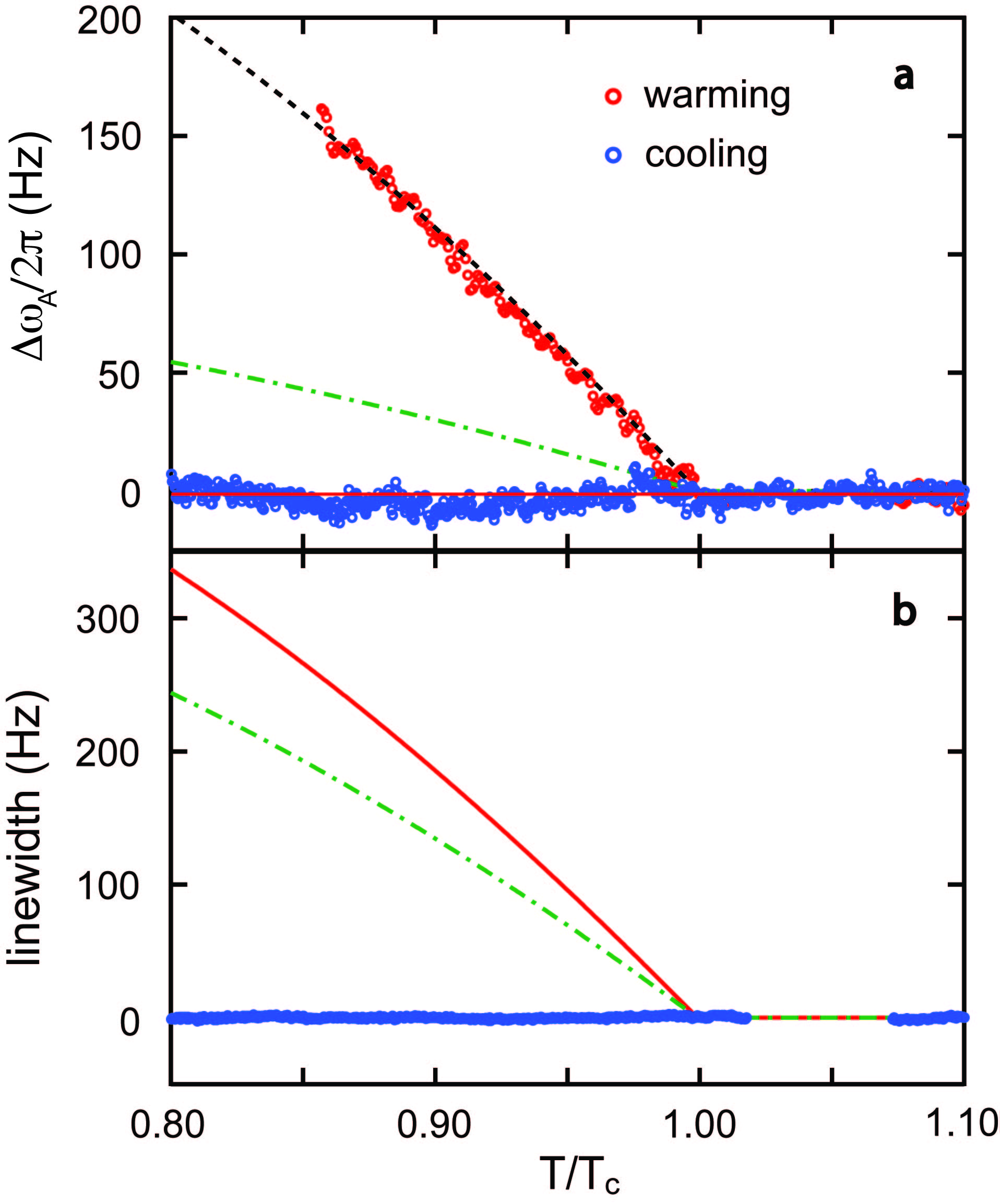}}%
\caption{\label{fig4}(Color online). {\bf{Identification of the glass phase.}} Comparison of spectra cooling and warming in $H=196$ mT with small NMR tip angle, $\beta = 8^{\circ}$ versus reduced temperature. a) Frequency shift on cooling (blue circles) from the normal state compared with warming (red circles) from the B-phase. The black dashed line is the frequency shift of the dipole-ordered axial state having an order parameter suppression of 25\,\%.  This suppression was inferred from the frequency shift measurements in the B-phase~\cite{Pol.11}  and the theoretically defined ratio of the shift of the isotropic state to the axial state for a  dipole-ordered configuration,  $\bm{\hat{l}}\,||\,\bm{\hat{d}}$, given by the Leggett relation~\cite{Leg.75} and Eq.~2, $\Omega_B^2/ \Omega_A^2 = (5/2) (\chi_A\Delta_B^2/\chi_B\Delta_A^2)$. The red solid (green dash-dot) curve  corresponds to the average frequency shift of a random 2D-disordered (3D-disordered) state. b) The additional contribution to the NMR linewidth beyond that of the normal state measured on cooling (blue circles). The linewidth data for  warming  are the same, Fig.~2(b), but are omitted for clarity.  The red solid (green dash-dot) curve is the  linewidth of a simulated random 2D-disordered (3D-disordered) state.}
\end{figure}

We use nuclear magnetic resonance (NMR) to look for the LIM state of superfluid  $^3$He-A, directly interrogating the orientation of $\bm{\hat{l}}$ by measuring the  Leggett  shift of the NMR spectrum, $\Delta \omega_A$.  In pure $^3$He this frequency shift  is proportional to the nuclear dipole energy, $F_D \propto - (\bm{\hat{l}} \cdot \bm{\hat{d}})^2$, where $\bm{\hat{d}}$ is  a spin-space vector constrained to be perpendicular to $\bm{\hat{s}}$ while minimizing $F_D$.  This shift is strongly temperature dependent~\cite{Leg.75}, but for an orbital glass it should be very small, or ideally zero (Supplementary Materials) as we report here.  

Leggett  interpreted a shift in the  NMR spectrum,  centered at the frequency $\omega_A(T)$, as evidence of  orientational order~\cite{Leg.75},
\begin{equation}
\omega_A^2(T) = \omega_L^2 + \Omega_A^2(T)
\end{equation}
\noindent
in the absence of any external influences on $\bm{ \hat{l}}$.  A global minimum of dipole energy corresponds to $(\bm{\hat{l}} || \bm{ \hat{d})}$ giving a maximal frequency shift at large magnetic fields $(H>30 $ G),
\begin{equation}
\Delta \omega_A \equiv \omega_A - \omega_L \approx  \Omega_A^2\left( T \right)/2\omega_L
\end{equation}
\noindent
where $\omega_L$ is the NMR frequency of the free atom.   The longitudinal resonance frequency, $\Omega_A$, is proportional to the amplitude of the maximum energy gap,  going to zero at the transition temperature, $T_c$, as is the case for all superconductors. In the presence of external fields such as aerogel induced disorder, the dipole energy and the frequency shift in Eq. 2 depend on the orientations of  $\bm{ \hat{l}}$ , $\bm{ \hat{d}}$ and $\bm{H}$ (Supplementary Materials).

Previous NMR studies of superfluid $^3$He in aerogel have resulted in a wide range of temperature dependent frequency shifts accompanied by significant broadening of the NMR  line suggestive of a distribution in $\Delta \omega_A$~\cite{Spr.95, Bar.00, Bau.04, Nak.07, Kun.07, Elb.08, Dmi.10}.  It is likely that  macroscopic inhomogeneity and anisotropy  in the aerogel influence the orientation of $\bm{ \hat{l}}$ (Supplementary Materials). To avoid this problem, we have grown highly homogeneous isotropic aerogel with a  98.2\,\% porosity which we have shown to be free of strain with an accuracy of $0.012\,\%$~\cite{Pol.08,Pol.11} using optical birefringence before and after the NMR experiments. The aerogel sample has a cylindrical shape as shown in Fig~1(c), with magnetic field, $\bm{H}$, along the vertical $z$-axis. Warming from our lowest temperatures we  identified  two superfluid aerogel phases~\cite{Pol.11}, first the B(isotropic)-phase and then the A(axial)-phase as shown in Fig.~1(d), familiar from pure $^3$He, but with a 25\,\% suppression of the order parameter amplitude at the pressure $P = 26$ bar. In the present  work we  perform NMR measurements  comparing warming and cooling following the horizontal black dashed line in Fig.~1(d) through all the superfluid transitions at $P = 26.3$ bar.

 The phase transitions from B-to-A on warming, and A-to-B on cooling,  are identified by a discontinuity in nuclear magnetic susceptibility, $\chi_{A(B)}$, characteristic of a first order thermodynamic transition, Fig.~2(a). The transitions are very sharp on both warming and cooling with transition width $\approx 0.2\%\,\,T_c$ indicative of a highly homogeneous sample~\cite{Pol.11} with typical NMR spectra shown in Fig.~2(b).  In Fig.~2(c) we provide an explicit demonstration that the A-phase  obtained on warming is indeed the axial superfluid state by measuring its  frequency shift as a function of  the NMR tip angle, $\beta$ (Methods Section),  and comparing with theory~\cite{Bri.75}.

The warming transitions shown in Fig.~3 have quadratic field dependence, as expected from  Ginzburg-Landau theory~\cite{Thu.98,Hal.08},  precisely mimicked by cooling transitions which supercool by $\sim 300\,\mu$K independent of magnetic field.   Extrapolation of the warming transition, $T_{BA}(H)$,  to zero field  gives $T_{BA}(0) -T_c = 0.8 \pm 16.3\,\mu$K indicating that for $H = 0$, the equilibrium state  is the isotropic B-phase.

In Fig.~4(a) we show our measurements  of the temperature dependence of the frequency shift of the A-phase, $\Delta \omega_A$ (Eq.~2), on cooling from the normal state (blue circles) compared with warming from the B-phase (red circles). This history dependence is unprecedented.  At first glance it might seem that  our observation of exactly zero shift on  cooling  corresponds to a strongly inhomogeneous distribution of frequency shifts that average to zero.  However, the NMR line shape is identical to the normal state, evident from the linewidth in Fig.~4(b),  ruling out an inhomogeneous  distribution of shifts.   Rather, we infer that cooling from the normal state generates a disordered, but spatially homogeneous, superfluid order parameter in the predicted  LIM state~\cite{Vol.08}.   In contrast, the A-phase produced by warming from the B-phase, Fig.~2(b),  has a uniform frequency shift corresponding to an axial state with the minimum possible dipole energy~\cite{Pol.11} shown by the black dashed curve in Fig.~4(a).  As in the case of pure  $^3$He, this behavior is a manifestation of long range orientational order. A possible reason for maximal  order in the  warming experiment can be associated with the presence of a phase boundary at this first order  transition,  orienting the angular momentum~\cite{Leg.75,Bra.08} and breaking rotational symmetry during the formation of the A-phase, thereby inhibiting a LIM state.  Another possibility is that the LIM state is disfavored at low temperatures, an explanation that will require  theoretical justification. In addition, it is worth noting that the superfluid transition from the normal state in aerogel is a second order thermodynamic transition and has no known hysteresis on cooling compared to warming \cite{Ben.11}. 

The direction of $\bm{\hat{l}}$ for $^3$He in the presence of aerogel should be locally uniform at least over a sufficiently small length scale $\xi_{LIM}$ (Fig. 1(a)) that depends on the degree of disorder; Volovik estimated this to be $\sim 1\,\,\mu$m (Reference \cite{Vol.08} and Supplementary Materials).  The direction of $\bm{\hat{d}}$ must be uniform over a distance, $\xi_D  \sim 8\,\,\mu$m (Fig.1(b))~\cite{Vol.90}, called a dipole length determined by the balance between dipole energy and gradient energy.  In the limit that $\xi_{LIM} < \xi_D$, $\bm{\hat{l}}$ is randomly oriented with respect to $\bm{\hat{d}}$.  As a result  the frequency shift collapses to zero from its equilibrium value expressed by Eq. 2 and the linewidth of the NMR spectrum will be identical to that of the normal state (Reference \cite{Vol.08} and Supplementary Materials).  In the other limit, $\xi_{LIM} > \xi_D$, $\bm{\hat{d}}$ can follow the projection of $\bm{\hat{l}}$ onto the plane perpendicular to $\bm{H}$.  Then the NMR spectrum will be inhomogeneously broadened and, in general, the average frequency shift will be non-zero. It has been established~\cite{Kun.07} that macroscopic strain in aerogel orients $\bm{\hat{l}}$ over length scales larger than $\xi_D$, so that inhomogeneity or anisotropy in partially strained samples~\cite{Elb.08,Dmi.10} might obscure the  LIM effect.


\begin{figure}
\centerline{\includegraphics[width=.35\textheight]{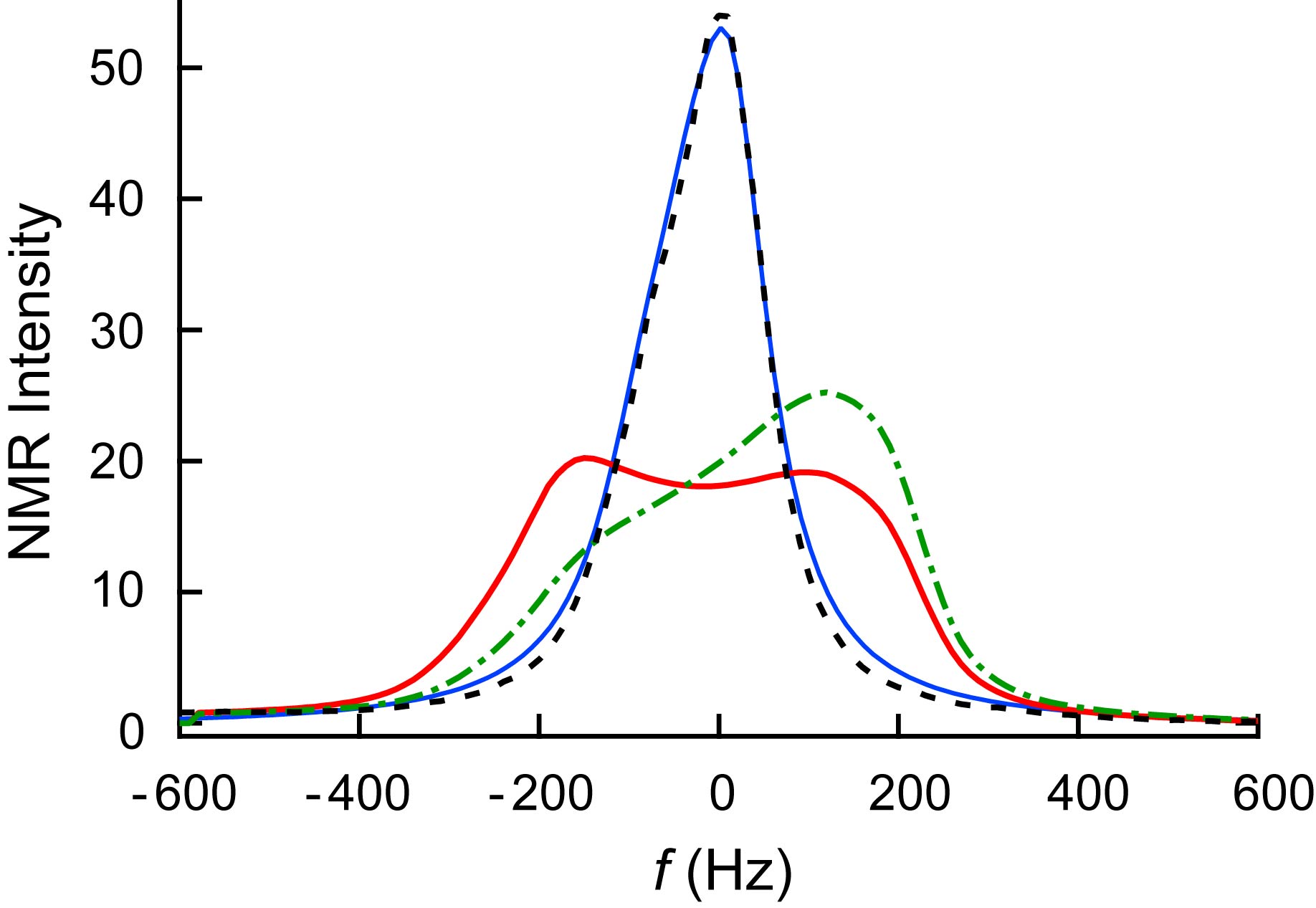}}  
\caption{\label{fig5}(Color online). {\bf{Comparing the glass phase with macroscopic inhomogeneity.}} Simulated NMR spectra for 2D-disordered and 3D-disordered distributions of $\bm{\hat{l}}$ compared to measured spectra in the normal state (dotted curve) and in the A-phase on cooling (blue curve) in $H=196$ mT at a temperature of $T/T_c =0.85$.  A LIM state spectrum should have no frequency shift nor linewidth broadening compared to the normal state spectrum, as is evident in our measured spectra.  However, the simulated 2D-disorder spectrum (red solid curve) and the 3D-disorder spectrum (green dash-dot curve), are significantly different, notably with an increase in linewidth in both cases and an increase in shift for 3D-disorder.}
\end{figure}

In the Larkin-Imry-Ma state the orbital angular momentum has no preferred direction down to and below the 10 $\mu$m scale given by the dipole length, $ \xi_D$, \emph{i.e.},  $\xi_{LIM} < \xi_D$.  As such it is very different from a macroscopically inhomogeneous state induced by  non-uniformity in density or anisotropy in the aerogel.  To illustrate the latter, we consider two models: $i)$ the 2D-disordered state where the orbital angular momentum is randomly disordered in a plane, \emph{e.g.}, the $xz$ or $yz$-plane, Fig.~1(c), and $ii)$ the 3D-disordered state where the orbital angular momentum is random in three dimensions. 

In order to compare our measurements with possible macroscopic distributions of angular momentum we have simulated  the NMR spectra for these two models by calculating the convolution of the normal state spectrum with  spectra corresponding to the probability distribution for the orientation of the order parameter in each model.  The result in Fig. 5 for the 3D-disordered state is the green  dash-dot curve; the red solid curve is for 2D-disorder.  We obtained the frequency shift from the first moment  of the simulated NMR spectrum and we calculated the linewidth from the second moment (Supplementary Materials).  In both cases we compared  results with our measurements in Fig. 4.

The 3D-disordered state has a positive shift that deviates from the data for both warming and cooling experiments (green dash-dot curve, Fig.~4). However the 2D-disordered state (red solid curve) has zero frequency shift, indistinguishable from our cooling measurements and that of the LIM state.  Based on frequency shift alone it is impossible to identify a LIM superfluid glass as distinct from a 2D macroscopically-disordered  distribution of the angular momentum.  On the other hand, the linewidths for both the 2D-disordered  and the 3D-disordered states, have significant increases according to our simulation as compared with a LIM state.  Our observations are inconsistent with models having significant macroscopic inhomogeneity supporting the conclusion that we have observed the 3D Larkin-Imry-Ma effect for superfluid $^3$He-A when it is cooled from the normal state.

In earlier NMR work on $^3$He in anisotropic aerogels, Elbs {\it et al.}~\cite{Elb.08} and Dmitriev {\it et al.}~\cite{Dmi.10} report evidence for a 2D-LIM effect. The spectra of Reference ~\cite{Dmi.10} for three different samples have some similarity to our model calculation for a 3D-macroscopically disordered state, i.e. they exhibit positive frequency shifts of various amounts for three different samples and significant linewidth broadening compared to the normal state. Nonetheless, Dmitriev \emph{et al.} performed NMR tip angle and magnetic field orientation experiments and reported that these are consistent with a 2D LIM state indicating that there is a LIM effect present, but superposed with effects of macroscopic inhomogeneity in the aerogel.  

Finally we note that in the absence of our ability to switch off the LIM effect we would not have been able to detect the presence of a superfluid until it appeared suddenly on cooling as a very  unusual first order transition to the B-phase. For sufficiently large magnetic fields such as to suppress the B-phase, evidence from NMR spectra for superfluid $^3$He would be completely hidden.


\subsection{Methods}
  The aerogel sample is the same as we used previously~\cite{Pol.11} having a cylindrical shape, Fig.~1(c), 4.0 mm in diameter and 5.1 mm long, with a measured porosity of 98.2\%. It was grown via the one-step sol-gel method~\cite{Pol.08} and characterized thoroughly with both optical-birefringence, cross-polarization techniques and, on similarly prepared samples, small angle X-ray scattering (SAXS)~\cite{Pol.11,Pol.08}.  We found the aerogel to be uniformly isotropic with resolution better than $20\,\mu \mathrm{m}^2$. We performed pulsed NMR experiments  at a pressure $P=26.3$ bar in a magnetic field range from $H=49.9$ to $196$ mT. The $H_1$ field that generates the RF pulse was oriented parallel to the cylindrical axis. The  RF pulse tips the nuclear magnetization by an angle $\beta$ away from the external field.  A Fourier transform  of the free induction decay (FID) signal of the magnetization in the time domain was  phase corrected to obtain the absorption spectrum. The magnetic susceptibility, $\chi$, was determined from the numerical integral of the phase corrected absorption spectrum.  Linewidths were calculated from the square-root of the relative second moment  of the spectrum. The sample was cooled via adiabatic demagnetization of PrNi$_5$ to a minimum temperature of 650 $\mu$K and NMR measurements with a constant small tip angle $\beta$ were performed while the sample warmed or cooled slowly through all the superfluid transitions, at varying rates of $\sim$ 3 to 10 $\mu$K/hr.  Thermometry was based on $^{195}$Pt NMR calibrated relative to the known  transition temperatures of pure $^3$He. 
  
\subsection{Acknowledgments}
We are grateful to  M.J.P. Gingras, J.A. Sauls, G. Volovik, V.V. Dmitriev, M.R. Eskildsen, and D. Vollhardt for helpful comments and for support from the National Science Foundation, DMR-1103625.

\end{document}


\title{Supplementary Information:\\
``The Superfluid Glass Phase of $^3$He\,-\,A''}

\author{J.I.A. Li}
\email[]{jiali2015@u.northwestern.edu}
\author{J. Pollanen}
\author{A.M. Zimmerman}
\author{C.A. Collett}
\author{W.J. Gannon}
\author{W.P. Halperin}
\email[]{w-halperin@northwestern.edu}
\affiliation{Northwestern University, Evanston, IL 60208, USA}

\date{\today}

\maketitle

\noindent
{\bf SI.1\,\,\, NMR response for LIM state}\\


The dipole energy for superfluid $^3$He-A  is~\cite{Vol.90},
\begin{equation}F_D=-\frac{1}{2}\Omega^2_A\frac{\chi_A}{\gamma^2}\left ( \bm{ \hat{l} \cdot \hat{d}}\right )^2.\label{SI1}\\\end{equation}
\noindent
The relative orientation of $\bm{\hat{l}}$ and $\bm{\hat{d}}$ can be parametrized by two angles $\theta$ and $\phi$, where $\theta$ is the angle between $\bm{\hat{l}}$ and the magnetic field $\bm{H}$; and $\phi$ is the angle between $\bm{\hat{d}}$ and the projection of $\bm{\hat{l}}$ in the plane perpendicular to the field. In our frequency range, $\omega_L \gg \Omega_A$, we can average the spin-orbit interaction over the fast Larmor precession ~\cite{Bri.75}, and over a volume the size of the dipole length, $\xi_D \sim 8\,\mu$m ~\cite{Vol.06},      \begin{equation}
      \bar{F}_D(\bm{r}) \equiv \frac{1}{2}\Omega^2_A\frac{\chi_A}{\gamma^2}\left \langle U(\bm{r}) \right \rangle. \hspace{120pt}\label{SI2}\\
      \end{equation}
      \noindent
     $\left\langle U(\bm{r}) \right \rangle$ can be expressed in terms of the NMR tip angle, $\beta$ (Methods Section),
          	\begin{eqnarray}
      \left \langle U(\bm{r}) \right \rangle=-\frac{1}{2} \sin^2\beta \hspace{140pt}\notag\\
      +\frac{1}{4}\left(1+\cos\beta \right)^2 \left \langle\sin^2\phi(\bm{r})\right \rangle \left \langle \sin^2 \theta(\bm{r}) \right \rangle \hspace{20pt}\notag\\
      -\left(\frac{7}{8}\cos^2\beta+\frac{1}{4}\cos\beta-\frac{1}{8} \right) \left \langle \sin^2 \theta(\bm{r}) \right \rangle. \hspace{20pt}\label{SI3}      
      \end{eqnarray} 
\noindent
Evidently, the dipole energy  is minimized when $\phi=0$ according to Eq.~SI3. However, if the orientation of $\bm{\hat{l}}$ changes on a length scale smaller than $\xi_D$, the cost in bending energy  is too much for $\bm{\hat{d}}$ to follow $\bm{\hat{l}}$, resulting in a non-zero value of $\phi$.  

\begin{figure}
\centerline{\includegraphics[height=0.35\textheight]{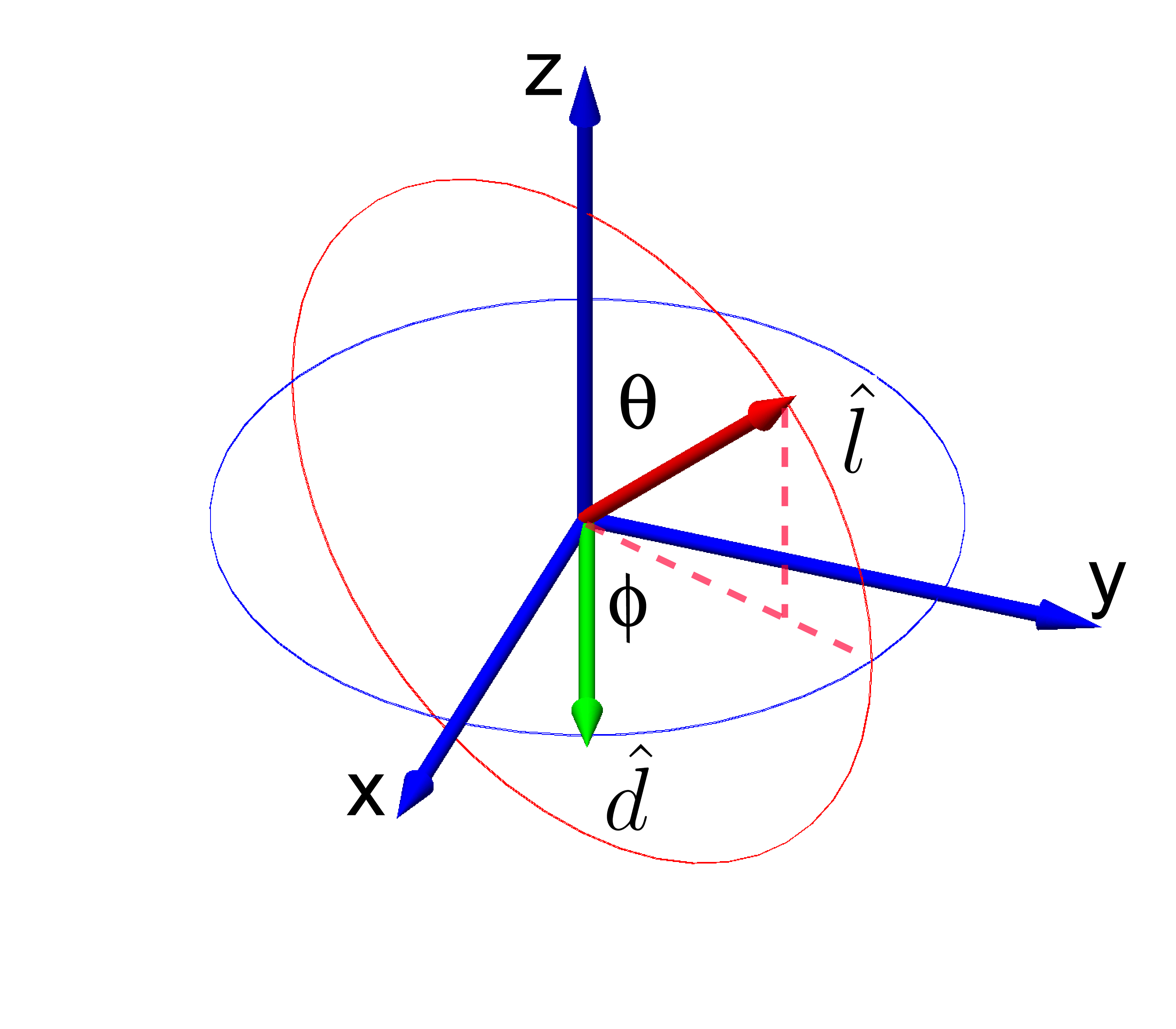}}
\caption{\label{fig1} (Color online). {\bf{Superfluid $^3$He\,-\,A order parameter.}} The relative orientation of $\bm{\hat{l}}$ and $\bm{\hat{d}}$. The magnetic field $\bm{H}$ is along the z axis, confining $\bm{\hat{d}}$ to be in the x-y plane. The angle between $\bm{H}$ and $\bm{\hat{l}}$ is $\theta$, while the angle between $\bm{\hat{d}}$ and the projection of $\bm{\hat{l}}$ in the x-y plane is $\phi$.}
\end{figure}

For complete microscopic disorder,  $\bm{\hat{l}}$ and $\bm{\hat{d}}$ take on all possible orientations with equal probability in a volume of the size of the dipole length $\xi_D$, {\it i.e.} $0 < \theta < \pi$ and $0 < \phi < 2\pi$, making $\left \langle \sin^2\theta \right \rangle$ and $\left \langle \sin^2\phi\right \rangle$ single-valued throughout the sample and easily shown to be,
     	\begin{eqnarray}
      \left \langle \sin^2 \theta \right \rangle = 2/3, \hspace{0pt} \label{SI4}\\
			\left \langle \sin^2   \phi \right \rangle = 1/2. \hspace{0pt} \label{SI5} 
       \end{eqnarray}
Since the frequency shift is~\cite{Vol.06},
      \begin{equation}
      \Delta\omega_A=\frac{\gamma}{\chi_A H}\frac{\partial \bar{F}_D}{\partial \cos\beta} =-\frac{\Omega^2_A}{2\omega_L}\frac{\partial \left \langle U \right \rangle}{\partial \cos\beta},\hspace{20pt}\label{SI6}\\
      \end{equation} 
\noindent
we can combine Eqs.~SI3 to SI6, to show that $\Delta\omega_A=0$ when $\beta=0$, meaning that a LIM state will have no frequency shift relative to the normal state in the small tip angle limit. Furthermore, since $\left \langle \sin^2\phi\right \rangle$ and $\left \langle \sin^2\theta \right \rangle$ are the same throughout the sample the LIM state does not contribute to the linewidth.  Consequently the NMR spectrum for the LIM state should be identical to that of the normal fluid exactly as we observe and as shown in Fig.~4. \\

\noindent
{\bf SI.2\,\,\, Calculation of the macroscopically disordered state spectra}\\

A disordered state with inhomogeneous textures of the order parameter induced by inhomogeneity or anisotropy in partially strained aerogel samples results in a distribution in the direction of $\bm{\hat{l}}$ relative to the external magnetic field, $\bm{H}$, on a length scale longer than $\xi_{D}$. In this case the angle between $\bm{\hat{d}}$ and the projection of $\bm{\hat{l}}$ in the plane perpendicular to the field, $\phi$, will be zero to minimize dipole energy, while the angle between the chiral axis and the magnetic field, $\theta=\acos({\bm{\hat{l}} \cdot \bm{H}})$, has a distribution $P(\theta)$ throughout the sample. This will lead to a distribution in the  dipolar fields and hence a distribution of frequency shifts in the NMR spectrum. For small tip angle, $\beta$, the frequency shift in the A-phase reduces to \cite{Bri.75,Vol.06},
\begin{equation}
		\Delta\omega_{A}(\theta)=-\frac{\Omega^{2}_{A}}{2\omega_{L}}\cos(2\theta).
		\label{SI7}
\end{equation}

Since the distribution of $\bm{\hat{l}}$ in $\theta$-space, $P(\theta)$, and the distribution in frequency space, $P(\omega)$ must satisfy $P(\theta)\delta\theta=P(\omega)\delta\omega$, we have:
\begin{equation}
		P(\omega)=P(\theta)\frac{\partial \theta}{\partial \omega}.
		\label{SI8}
\end{equation}

For a random 2D-disordered state, $\bm{\hat{l}}$ is disordered randomly in the $xz$ or $yz$-plane shown in Fig.~1a, then, $P_{2D}(\theta)=1$. On the other hand a random 3D-disordered state leads to $P_{3D}(\theta)=\sin \theta$. Combining Eq.~SI7 and Eq.~SI8,

\begin{eqnarray}
     P_{2D}(\omega)&=&\frac{1}{\sqrt{1-\left (\frac{\omega}{\omega_M}  \right )^2}},\hspace{30pt} \left | \omega \right | \leq \omega_M, \label{SI9}\\
     P_{3D}(\omega)&=&\frac{1}{\sqrt{1-\frac{\omega}{\omega_M}}}, \hspace{50pt}\left | \omega \right | \leq \omega_M,\label{SI10}
\end{eqnarray}
where $\omega_M = \frac{\Omega^{2}_{A}}{2\omega_{L}}$. If the normal state lineshape is given by $F_{n}(\omega-\omega_{L})$, then the lineshape of the disordered state is given by the convolution product of the normal state line and $P(\omega)$

	\begin{equation}
	F_{D}(\omega)=\int \! P(\omega^{'})F_{n}\left(\omega-\omega_{L}-\omega^{'}\right) \,d\omega^{'}.
	\label{SI11}
	\end{equation}
Since the normal state spectrum is obtained experimentally and shown in Fig.~5 (dotted curve), the 2D and 3D-disordered spectra are simulated by combining Eq.~SI9, SI10 and SI11, and shown in Fig.~5 as red (solid) and green (dash-dot) curves. \\

\noindent
{\bf SI.3\,\,\, Volovik's theory  of the Larkin-Imry-Ma effect}\\

Here we reproduce Volovik's arguments\cite{Vol.06,Vol.08} for the Larkin-Imry-Ma effect for $^3$He-A in aerogel. 

According to the theory of Rainer and Vuorio,\cite{Rai.77} for a cylinder  with diameter $\delta$ smaller than the coherence length $\xi_0$  immersed in $^3$He-A, the $\bm{\hat{l}}$-vector remains uniform outside the  cylinder  at a cost of orientational energy:

\begin{equation}
		E_i=E_a \left( \bm{\hat{l}} \cdot \bm{\hat{n}} \right)^2 , \,\,\,\,\,\,\,\,E_a \sim \frac{\Delta^2}{T}{k_F}^2 \xi_a \delta
		\label{SI12}
\end{equation}
where $\bm{\hat{n}}$ is the direction of the cylinder axis, $k_F$ is the Fermi momentum, $\Delta$ is the superfluid energy gap while $T_c$ is the superfluid transition temperature. The parameter $E_a$ is positive as follows from Ref.~\cite{Rai.77}. Obviously, it is energetically favorable to have the $\bm{\hat{l}}$-vector aligned perpendicular to the cylinder axis.

In Volovik's theory of the Larkin-Imry-Ma effect \cite{Vol.08}, it is assumed that aerogel consists of randomly oriented cylindrical strands with diameter $\delta \sim 3$ nm, while the length of the strand and the distance between different strands are both $\xi_a \sim 20$ nm. Since the diameter of the cylinders is much smaller than $\xi_0$, the theory of Ref.~\cite{Rai.77} can be applied.  If the alignment of the strands is  random, meaning the aerogel is macroscopically homogeneous and isotropic, the orientational effect of the randomly aligned strands is spatially averaged to zero. However, local fluctuations of the strand orientation will result in a locally preferred $\bm{\hat{l}}$-vector orientation. Consider a box of size $L^3$, which contains $N = \frac{L^3}{\xi_a^3} \gg 1$  randomly oriented cylinders labelled $i$.  Due to the fluctuation of orientational energy, the energy gain for having the $\bm{\hat{l}}$-vector aligned in the same direction in this box was estimated\cite{Vol.08} to be:
\begin{equation}
		E_{random}=-\left \langle \sum_{i=1}^{N} \left ( E_i - \left \langle E\right \rangle\right )^2\right \rangle^{\frac{1}{2}} \sim -E_a \xi_a^{-\frac{3}{2}}L^{-\frac{3}{2}}.
		\label{SI13}
\end{equation} 
The neighboring boxes have different $\bm{\hat{l}}$-vector orientation, resulting in a gradient energy:
\begin{equation} 
  E_{grad} \sim K \left( \triangledown \bm{\hat{l}}\right)^2 \sim \frac{K}{L^2}  
		\label{SI14}
\end{equation} 
where K is the rigidity of $\bm{\hat{l}}$, which can be estimated as $K \sim \left( \frac{k_F^3}{m}\right)\left(\frac{\Delta^2}{{T_c}^2} \right)$. Since the random energy competes with the gradient energy, the optimal size of the box, $\xi_{LIM}$, can be achieved by minimizing the combination of Eq.~SI13 and ~SI14:
\begin{equation}
		\xi_{LIM} \sim \frac{K^2 \xi_a^3}{E_a^2} \sim \xi_a \frac{\xi_0^2}{\delta^2}.
		\label{SI15}
\end{equation} 
With $\xi_0 \sim \xi_a \sim 20$ nm and $\delta \sim 3$ nm, one finds $\xi_{LIM} \sim 1 \mu$m, which is the length scale over which the long range orientational order (LROO) is destroyed by randomly oriented aerogel strands. In this model, $\xi_{LIM} \sim 1 \mu$m $\ll \xi_D \sim 8 \mu$m.  On all length scales greater than a dipole length $\xi_D$, $\bm{\hat{l}}$ takes on all possible orientations and the gradient energy makes it energetically more favorable to decouple the $\bm{\hat{l}}$ and $\bm{\hat{d}}$-textures. As a result $0 < \theta < \pi/2$ and $0 < \phi < 2\pi$ within any volume the size of $\xi_D$, and it follows that $\left \langle \sin^2\phi\right \rangle$ and $\left \langle \sin^2\theta \right \rangle$ are both single valued throughout the sample.